\documentclass[prl,aps,showpacs,twocolumn,floats]{revtex4} \input epsf
\usepackage{bm}

\usepackage{amsmath}

\usepackage{graphicx}
\usepackage{dcolumn}
\usepackage{bm,color,soul,ulem}
\usepackage{url,hyperref,varioref}
\hypersetup{colorlinks,citecolor=blue,linkcolor=red,urlcolor=blue}

\begin{document}
\title{Spontaneous tubulation of membrane vesicles coated with bio-active filaments.  }
\author{$^1$ Gaurav Kumar}
\author{$^2$ N. Ramakrishnan}
\author{$^1$ Anirban Sain}
\email{asain@phy.iitb.ac.in}
\affiliation{$^1$ Physics Department, Indian Institute of  Technology-Bombay, 
Powai, Mumbai, 400076, India. $^2$ Department of Bioengineering, University of Pennsylvania, Philadelphia, PA, 19104, USA}

\begin{abstract}
{Narrow membrane tubes are commonly pulled out from the surface of  phospholipid vesicles using forces 
applied either  through laser or magnetic tweezers or through the action of processive motor proteins. 
Recent examples have emerged where array of such tubes spontaneously grow from vesicles coated with 
bioactive cytoskeletal filaments (e.g. FtsZ, microtubule) in the presence GTP/ATP. We show how a soft 
vesicle deforms as a result of the interplay between its topology, local curvature and the forces due 
to filament bundles.  We present results from Dynamically Triangulated Monte Carlo simulations of a 
spherical continuum membrane  coated with a nematic field (the filaments) and show how the intrinsic 
curvature of the 
filaments and their bundling interactions drive membrane tubulation. We predict interesting patterns 
consisting of large number of nematic defects which accompany tubulation. A common theme emerges 
that defect locations on vesicle surfaces are hot spots of membrane deformation activity, which could 
be useful for vesicle origami. Although our equilibrium model is not applicable to the  nonequlibrium 
shape dynamics exhibited by active microtubule coated vesicles, we show that some the features like 
size dependent vesicle shape 
can still be understood from our equilibrium model. }
{E-mail: asain@phy.iitb.ac.in}
\end{abstract}
\maketitle

\section{Introduction}
Narrow membrane tubes are ubiquitous in eukaryotic cells and are essential for a number of cell functions including signalling and trafficking. 
Examples of tubular structures are the axons and dendrites of nerve cells, tubular networks in the Golgi body and the Endoplasmic reticulum. The curvature energy of a tubular protrusion $E \propto \kappa (l/r)$, where $l$ is the length of the tube, $r$ the radius and $\kappa$ the membrane bending rigidity. As a result the formation of narrow tubes (characterised by $l>>r$) requires
 extremely high energies and hence they are not expected to be stable unless stabilised by external forces \cite{julichertube}.  
 Such forces can be applied by laser tweezers \cite{lasertweezer}, suction pressure \cite{micropipette} or processive molecular motors. \cite{motor}. 
 Forces internal to the cell or membrane vesicles, for example, growing bundles of actin \cite{actinmembranePlos,MesarecIglic2017} or microtubule (MT) filaments \cite{libchaberbuckledMT,Bausch} can also push out  membrane protrusions which can grow into long tubes upon further polymerization
of the filaments. 
 Tubulation can also be promoted by spiral shaped protein filaments like dynamin 
\cite{dynamin,hinshaw2000dynamin} and ESCRT-III \cite{ESCRT3}, or proteins with curved domains
like Bar \cite{peter2004bar}, ENTH \cite{ford2002curvature}, Exo70 \cite{zhao2013exo70}, etc that wrap around the tube. Such induced 
membrane curvatures are interpreted as local  spontaneous curvature \cite{kozlov}. Spontaneous membrane curvature  can  also naturally result from lipid heterogenity 
\cite{Dimova} or lipid tilt \cite{lipidtilt} in a phospholipid membrane.


\begin{figure}[h]
\centering
\includegraphics[width=3in]{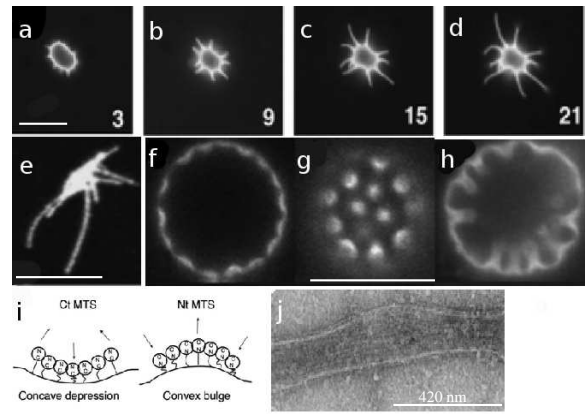} 
\caption {(color online) Tubulation of a vesicle, (a) to (d), due to FtsZ filaments (fluorescent regions),  in real time (minutes).
 (e) shows loss of spherical shape  due to excessive tubulation. (f) and (g) show high resolution images focused at the   equatorial plane and the surface, respectively, of the tubulated vesicle.  Nearly regular pattern of tubulation sites are visible. (h) shows reverse tubulation i.e., convex bulges on the vesicle surface accompanied by membrane invaginations protruding inward. (i) shows two different ways of anchoring FtsZ onto the membrane, causing opposite membrane curvatures. (j) shows parallel arrangement of FtsZ filaments along the tubular axis. Scale bars are $10\mu$m in the upper and middle rows, and $420nm$ in the lower row. Adapted from Fig.1,2,3 and 6 of Ref\cite{osawaEMBO}, with permission.}
\label{fig.expt}
\end{figure}


Here we address experiments \cite{osawaEMBO} where an array of membrane tubes emerge spontaneously due to surface active filamentous proteins FtsZ, attached to the outer surface of artificial membrane vesicles (liposomes). Interestingly, here the FtsZ filaments align with the tube axis, instead of wrapping around the tube. We attribute this to the
attractive interaction among parallel FtsZ filaments, in the presence of GTP, which gives rise 
to FtsZ bundles \cite{bundle}.  In our model, this bundling interaction will be the main driving force towards multiple tube formation.
Secondly, the regular arrangement of many tubes on the curved vesicle surface 
is rather striking and has not been modelled before.  As we will see later that the formation of such tubes cannot be reliably modelled by considering just a single axisymmetric tube because collective effects turn out to be important. For example, 
in the present case of tubular array the boundary condition at the base of a single tube is not axisymmetric but instead has five or six-fold discrete 
rotational symmetry depending on the number of neighbouring tubes surrounding it.

In this experiment Osawa et. al.~\cite{osawaEMBO} attached membrane targeted FtsZ filaments onto the outer surface of large Giant Unilamellar Vesicles (GUVs) of diameter $\sim 10\mu$m and observed tubes of diameter $50-200$nm grow spontaneously.  
The filaments caused either concave depressions or convex bulges on the membrane, as illustrated in Fig.~\ref{fig.expt}(i), 
depending on whether the membrane anchors  (FtsA) were attached to the C terminal or the N terminal of the filaments, respectively. Depending on the type of attachment, membrane tubes  either grew outward (see  time series in Fig.~\ref{fig.expt}(a-d)) or inward (see Fig.~\ref{fig.expt}h).  While elongation of tubes was observed only when GTP was  in abundance, tubes shrunk when medium was depleted of GTP, implying the essential role played by GTP.  But unlike in active systems here GTP hydrolysis neither generate any active forces
nor move the FtsZ filaments. In fact, here the FtsZ filaments are anchored to the membrane surface and they can at the most diffuse slowly.

In order to understand the FtsZ assisted membrane tubulation it is necessary to review the well known
physical properties of FtsZ filaments. FtsZ filaments have intrinsic curvature of order 
($100$-$200$ nm)$^{-1}$ and they play important role during cytokinesis of rod shaped bacteria 
like E. coli and B. subtilis \cite{ Sci17Huang,Sci17garner}. 
In the presence of GTP, FtsZ filaments also condense into bundles via weak, 
lateral,  inter-filament attraction \cite{zringprl}. These bundles can also locally bend membrane 
and generate constriction forces (few pN) \cite{osawaSci} on 
relatively wide membrane tubes of diameter $1-2\mu$m. 
Given these properties it has been unclear how these tubes form. 
In this article, treating the filaments as a nematic field  and using a generalised 
Canham-Helfrich \cite{canham70,helfrich73, SunilPRE,Sunilbiopj,SunilMacromolecule} 
model for the lipid membrane, we suggest a mechanism for vesicle tubulation and predict the 
pattern of arrangement of FtsZ filaments on the vesicle. 


According to the Hairy-Ball theorem \cite{prost} filaments, approximated as nematics here, 
cannot be arranged on a closed surface (tangentially) without
forming topological defects and the topological charge of these defects must add up to two. 
Arrangement of these defects is also a theoretically well studied 
problem \cite{prost,nelson,bowickPRL}. 
Keber et al \cite{Bausch} has recently studied active 
dynamics of such defects on spherical as well as deformed vesicles. However the influence of 
the nematics (filaments) on the elastic membrane is also very interesting because 
it leads to nontrivial deformation of the vesicle shape \cite{osawaEMBO,Bausch}, which is 
the focus of this article. In fact, in the case of FtsZ shape deformation of the vesicle 
is accompanied by large number of high energy defects which was not considered before. 

Qualitatively, Fig.~\ref{fig.expt}b-d,f suggest that concave depressions and tubes go hand in hand 
and we can guess that tubes may form at the junctions where such concave patches meet. 
The question then arises, how will these concave patches arrange themselves on a spherical surface.
Fig.~\ref{fig.expt}g further shows a regular arrangement of bright patches, with coordination number 
five or six. 
Can we then approximate the patches as the pentagons and hexagons that cover the surface of
a soccer ball ? The tubes would then emerge from the vertices. We will find out later that 
a variant of this picture is correct.

\section{Model} 
In our coarse-grained approach, we model the FtsZ coated membrane as a nematic field 
adhering to a deformable fluid membrane surface. This model was developed and many of its 
properties were studied by one of the authors \cite{SunilPRE,Sunilbiopj,SunilMacromolecule}. 
We will later highlight the results that are new here. In this model the local orientation 
of the nematic  field is denoted by the unit vector $\hat n (\vec r)$ which lies in the local 
tangent plane of the membrane and is free to rotate in this plane. Filament-membrane 
interactions are modelled as anisotropic spontaneous curvatures of the membrane, in the 
vicinity of the filament, while filament-filament interactions are modelled by the splay and 
bend terms of the Frank's free energy for nematic liquid crystals.  The total energy is  
\begin{eqnarray}
  E&=& \int dA \Big [\frac{\kappa}{2}  (2H)^2 + \frac{\kappa_{\parallel}}{2}(H_{\parallel}-c_{\parallel})^2 +
\frac{\kappa_{\perp}}{2}(H_{\perp}-c_{\perp})^2 \Big ]\nonumber\\ 
&+& \int dA \Big [\frac{K_1}{2} (\tilde\nabla.\hat n)^2 + \frac{K_3}{2}(\tilde\nabla. \hat t)^2\Big ].
\label{eqn:totalE}
\end{eqnarray}

Here the first term is the Canham-Helfrich elastic energy for membranes~\cite{canham70,helfrich73} with bending rigidity  $\kappa$   and membrane mean curvature $H=(c_1+c_2)/2$. Here, $c_1$ and $c_2$ are the local principal curvatures on the membrane surface along orthogonal tangent vectors $\hat t_1$ and $\hat t_2$.   $\kappa_{\parallel}$ and $\kappa_{\perp}$ are the induced membrane bending rigidities  and  $c_{\parallel}$ and $c_{\perp}$ are the induced intrinsic curvatures along $\hat n$, the  orientation of the filament in the local tangent plane,  and $\hat t$, its perpendicular direction, respectively. Origin of a nonzero $c_{\perp}$, an induced curvature 
perpendicular to the filament, is not obvious. In fact it will turn out to be the driving force towards 
filament bundling which accompanies membrane tubulation. Tubulation can also occur due to 
the $c{\parallel}$ term alone, however it does not promote formation of long straight filament bundles. 
The filament orientations on the tubes are different in these two cases. 

The membrane curvature along $\hat n$ and $\hat t$ are given by $H_{\parallel}=c_1 \cos^2 \phi + c_2 \sin^2 \phi$ and $H_{\perp}=c_1 \sin^2 \phi + c_2 \cos^2 \phi$, where $\phi$ denotes the angle between filament orientation $\hat n$ and principal direction  $\hat t_1$. $K_1$ and $K_3$ are the splay and bend elastic constants for the in plane nematic interactions and $\tilde \nabla$ is the covariant derivative on the curved surface~\cite{Chaikin}.

As in Ref~\cite{SunilPRE}, we use a discrete form of this energy functional to perform Monte-Carlo 
simulations on a triangulated membrane, 
to study the equilibrium shapes.
Each vertex $i$ hosts a orientation vector $\hat n_i$. In particular, we use the standard Lebwohl-Lasher model $E_{nn}=-\epsilon_{LL}\sum _{i>j}(\frac{3}{2} (\hat n_i.\hat n_j)^2 - \frac{1}{2})$\;, \cite{LL} to mimic the in plane nematic interaction terms. 
Here $\epsilon_{LL}$ is strength of the nematic interaction,  in one constant  approximation( $K_1=K_3$ ), and the sum 
$\sum _{i>j}$ is over all the nearest neighbour $(i,j)$ vertices on the triangulated grid, promoting alignment among the 
neighbouring orientation vectors. 

Models with anisotropic membrane curvatures have been developed and used by various authors
\cite{iglic2005,AnisotropicMemCurv2016simul,iglic2018}. But all of these efforts focussed on membrane 
structures that are axis-symmeric in nature and mainly sought to study formation of single tubes. 
Previous work by one of the authors \cite{SunilPRE,Sunilbiopj,SunilMacromolecule} focussed on effect of nonzero 
$\kappa_{\parallel}$ and $c_{\parallel}$. This amounts to introducing one preferred length 
scale, namely the radius of the tube. In contrast, the present work focusses on the effects of non-zero 
$\kappa_{\perp}$ and $c_{\perp}$. In fact, using these parameters we account for a new physical effect, namely 
bundling of filaments due to inter-filament attraction. 

Nonzero values of $\kappa_{\perp}$ and $c_{\perp}$ induce parallel alignment of filaments, on 
the outer surface of a membrane tube, all pointing along the tube axis ($\hat z$). Higher the 
$c_{\perp}$  smaller is the tube radius; this is equivalent to filament bundling. In our simulation, 
with fixed number of vertices, the density of vertices goes up relatively at 
the high curvature regions of the triangulated surface. This makes the number density of
vertices relatively higher at the tube surfaces, leading to higher filament density 
on the tubes, consistent with filament bundle picture. Since our model does not distinguish 
between the outer and the inner surfaces of a tube, this effect can also account for MT
bundles which are responsible for pushing out membrane tubes from a vesicle. 
Note that, the nematic interaction term in our model is also minimized when the filaments align 
with the tube axis ($\hat z$). This is also true for continuum Frank's free-energy where both 
splay the bend terms are minimized. But the nematic interaction term does not control
the tube radius because it is not dependent on the curvature of the underlying membrane 
as long as all the nematics on the tube are aligned along the common $\hat z$ axis of the straight tube.  
  
We will later show that bundling assisted membrane tubulation is possible even in the absence of intrinsic 
filament curvature (i.e., $c_{\parallel}=0$). This will be relevant for MT filaments which does not have 
any intrinsic curvature.  But in the case of FtsZ, since intrinsic curvature and bundling both are known to be 
involved, we will use nonzero values for both $c_{\parallel}$ and  $c_{\perp}$ for modeling FtsZ. 

\section{Results}
\begin{figure}[h]
\centering
\includegraphics[width=3.5in]{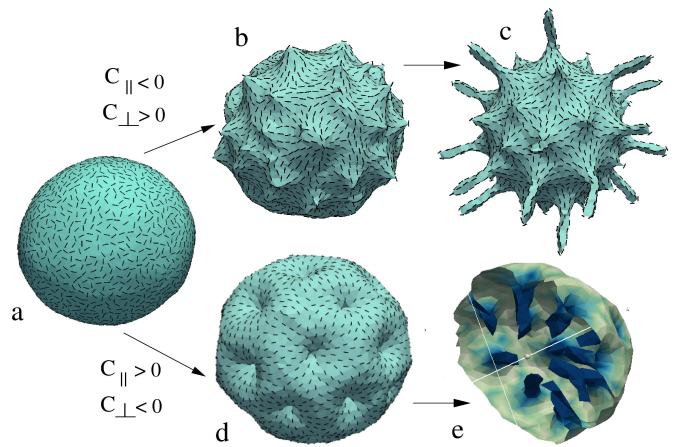}
\caption{ (color online) Growth of tubes, outward (a,b,c) or inward (a,d,c), depend on the signs of 
$c_{\parallel}$ and $c_{\perp}$. We use  $c_{\parallel}=-0.05$ and $ c_{\perp}=1$ for 
outward growing tubes, and reverse the signs of $c_{\parallel}$ and $c_{\perp}$ 
to get inward growing tubes. Other parameters are $\kappa_{\parallel}=35,\kappa_{\perp}=25$
and $\epsilon _{LL}=3$, in units of $k_BT$. Volume has been held fixed. 
In (e) the vesicle is sectioned to make the inner tubes visible 
The results from constant area ensemble are nearly identical at same parameter values. 
}
\label{fig.tubes}
\end{figure}

Monte-Carlo simulations of our model (Eqn.~\eqref{eqn:totalE}) show that, for $c_{\parallel}<0, c_{\perp}>0$,
with $|c_{\parallel}|<< |c_{\perp}|$, regularly spaced narrow tubes emerge (see Fig.~\ref{fig.tubes}) from 
an initially  spherical vesicle. We checked that $c_{\perp}$ mainly controls the tube radii and
$c_{\parallel}$ controls the curvature of the valleys. However the total number of tubes is a joint effect
of both $c_{\perp}$ and $c_{\parallel}$.
Reversing the signs of the intrinsic curvatures, i.e.,
$c_{\parallel}>0, c_{\perp}<0$ tubes grow into the vesicle (see Fig.~\ref{fig.tubes} a,d,e).  
Since  FtsZ coated membrane is  expected to be stiffer than the bare bilayer membrane (for which  $\kappa\sim 20k_BT$, 
we set $\kappa_{\parallel}=35k_BT$ and $\kappa_{\perp}=25k_BT$.
We found that setting $\kappa=0$ or $\kappa=20k_BT$ only makes minor qualitative 
differences in the tubular structures; tubes are thicker and little less in number with 
$\kappa=20k_BT$. Furthermore, nonzero $\kappa$ offers an initial energy
barrier for tube nucleation making tubulation slow, which could be bypassed by
raising the temperature temporarily in our Monte-Carlo simulation. 

\begin{figure}[h]
\centering
\includegraphics[width=3in]{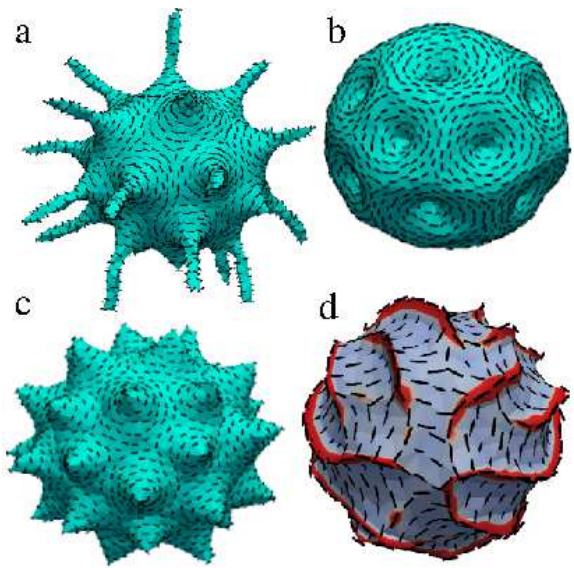}
\caption{ (color online) Shapes with fixed $c_{\perp}=0,\kappa_{\parallel}=25$ and $\epsilon_{LL}=5.0$, 
with other parameters varying, a) $\kappa=0,\kappa_{\perp}=25,c_{\parallel}=1.0$ b) same 
as (a) except $c_{\parallel}=-1.0$, c) $\kappa=10,\kappa_{\perp}=0,c_{\parallel}=1.0$ and 
d) same as (c) except $c_{\parallel}=-1.0$. }
\label{fig.4shapes}
\end{figure}

In Fig.\ref{fig.4shapes} we show few shapes, some of them quite unexpected ones, 
when the  bundling effect is switched off i.e., $\kappa_{\perp}=0$.
Yet, tubes or inverted tubes can form (see Fig.\ref{fig.4shapes}a and b, respectively)
provided $|c_{\parallel}|\gg 1/R$, the curvature of the original spherical vesicle.
However,   Fig.\ref{fig.4shapes}c and d, are examples where tubes do not form due to
variation of other parameters.
The criteria for emergence of a tube and dependence of its radius ($r$) on various parameters 
of the model can be understood by the stability analysis of a single tube given in 
Ref\cite{Sunilbiopj}. However the specific arrangement of many tubes on the vesicle surface 
cannot be inferred analytically. Minimization of the free energy of a straight, uniform 
cylindrical tube \cite{Sunilbiopj} yields equilibrium tube radius 
\begin{equation}
r^2=\frac {\frac{\kappa}{2}(\kappa_{\parallel}+\kappa_{\perp}) + 
\kappa_{\parallel}\kappa_{\perp}} {\kappa_{\parallel}\kappa_{\perp} 
(c_{\parallel}+c_{\perp})^2}
\end{equation}
and inclination ($\phi$) of the nematic \cite{Sunilbiopj} on the surface of the tube,
\begin{equation}
\cos^2 \phi = \frac{(\kappa_{\parallel}c_{\parallel} - \kappa_{\perp} c_{\perp}) r + 
\kappa_{\perp} }  {\kappa_{\parallel} + \kappa_{\perp}} 
\label{Eq.cosphi}
\end{equation}

Note that, if any of $\kappa_{\parallel}$ or $\kappa_{\perp}$ is 
zero, while $\kappa\neq 0$, the radius becomes infinite, siganalling that tubes will not 
form, which is the case for Fig.\ref{fig.4shapes}-c,d.  On the other hand if  
$\kappa=0$ then the tube radius is $(c_{\parallel}+c_{\perp})^{-1}$. This is the case for 
Fig.\ref{fig.4shapes}-a,b, where only $c_{\parallel}$ is nonzero and same for 
Fig.\ref{fig.tubes} where $c_{\perp}$ dominates over $c_{\parallel}$.
In this respect we note that in Ref\cite{SunilPRE}, Fig.11 despite $\kappa_{\perp}=0$,
and $\kappa,\kappa_{\parallel}\neq 0$, tubes still emerged. But these tubes were bent 
and also had nonuniform cross-section which violate the assumption in the stability analysis.
Furthermore, Eq.\ref{Eq.cosphi} suggests that for physically acceptable solutions the right
hand side (r.h.s.) must be between zero and one. Indeed $\phi=\pi/2$ for Fig.\ref{fig.tubes},  
while $\phi=0$, for Fig.\ref{fig.4shapes}a, consistent with the longitudinal and azimuthal orientation
of filaments on the tube, respectively. But when the r.h.s. is less than zero a boundary 
minima occurs for the free energy at $\phi=\pi/2$ and similarly for the r.h.s larger than 
one the free energy at $\phi=0$ is the physically acceptable minimum value. Correspondingly,
the formula for equilibrium $r$ also changes (see Ref\cite{Sunilbiopj}).

\begin{figure}[h]
\centering
\includegraphics[width=2in]{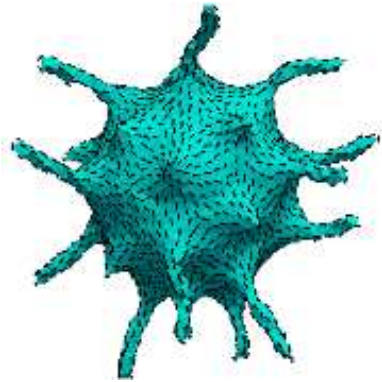}
\caption{ (color online) Tubulation even without intrinsic filament curvature, i.e., $c_{\parallel}=0$.
Other parameters are $\kappa=0,\kappa_{\parallel}=\kappa_{\perp}=25, c_{\perp}=1$ and $\epsilon_{LL}=5.0$ }
\label{fig.onlykper}
\end{figure}

Fig.\ref{fig.onlykper} shows that, even without any intrinsic curvature (i.e., $c_{\parallel}=0$)
of the adhering filaments, tubes may still emerge due to bundling interaction among filaments.
As we discuss later this may be the driving force for tubulation for vesicles coated
with MT, which does not have any intrinsic curvature like FtsZ.     

In previous work on this model \cite{SunilPRE,Sunilbiopj,SunilMacromolecule} neither volume nor area of 
the deformed vesicle was conserved during Monte-Carlo simulation. In this work we ran separate simulations
for fixed volume and fixed area ensembles.  We noticed no significant change in the qualitative results between
these two  different ensembles, except that  the simulation with constant area was relatively faster.  Experimentally, 
total fluorescence of the membrane bound FtsZ was found to increase during tubulation 
\cite{osawaEMBO}. This could be due to addition of higher density of filaments to the 
tubes from the solution or could be due to release of entropic membrane folds leading 
to increase in vesicle area, attracting more FtsZ from the solution. 


In Osawa's experiments \cite{osawaEMBO} tubes grew indefinitely (see Fig~\ref{fig.expt}e)
as long as GTP was supplied and they shrunk when GTP was depleted. In our simulation too
tube growth did not stop when the vesicle volume was kept fixed but the area was unconstrained.
GTP depletion is known to cause two things: a) it switches off the attraction between FtsZ filaments
and as a result they unbundle  \cite{bundle}, and b) it raises the intrinsic curvature of the FtsZ 
filaments \cite{osawaEMBO}. We  implement GTP depletion by turning off $\kappa_{\perp}$ to zero which causes 
the tubes to shrink. We also raise the value of $c_{\parallel}$ simultaneously, which marginally
increases undulations on the nearly spherical vesicle.

It can be argued that in a Monte-Carlo simulation only the final equilibrium state has physical 
relevance, while the intermediate states may not follow the actual system kinetics. Therefore 
we considered ensembles where both the volume and the area of the vesicle were fixed. 
This is a typical recipe for simulating, for example, red blood cell shapes \cite{wortis}. 
We fixed the area to be about $10\%$ excess over that of the corresponding sphere at a given volume. 
This recipe is physically meaningful for our FtsZ case also because there is a limit to the 
maximum amount of area that a vesicle can reserve in the form of membrane folds, beyond which 
area stretching elasticity becomes important.  For this ensemble, our system took the same 
pathway as before (when area was not fixed) but the tubulation was not indefinite and the 
system reached equilibrium, as in Fig~\ref{fig.expt}d,f 
or g. This ensemble becomes particularly important for the MT induced tubulation of deflated 
membrane vesicles in the experiments of Keber et al \cite{Bausch}, which we discuss now.

Keber et. al.~\cite{Bausch} attempted to mimic the active cell cortex by assembling 
arrays of  microtubules  on the inner surface of a spherical membrane vesicle, along with 
high concentration of kinesin motors. This rendered activity to the MT layer. MT filaments
showed incessant growth, shrinkage, bundling and sliding motion at the spherical surface.
Four topological nematic defects of charge +1/2 formed at the surface and showed interesting 
spatio-temporal dynamics. Upon partial deflation these vesicles deformed into ellipsoidal 
shape and four narrow tubes emerged (Fig.6c), with MT bundles inside. The tubes kept changing their positions.
Furthermore, for smaller vesicles only two tubes persisted (Fig.6d) along with shape changing dynamics. 
Few groups \cite{onsphere,voigt}, including Keber et al \cite{Bausch}, modelled 
the active MT-membrane system as nematics on a spherical surface and studied the 
effective dynamics of the nematic defects.  But so far no study has included vesicle 
deformation due to the active MT-motor dynamics.

\begin{figure}[h]
\centering
\includegraphics[width=3in]{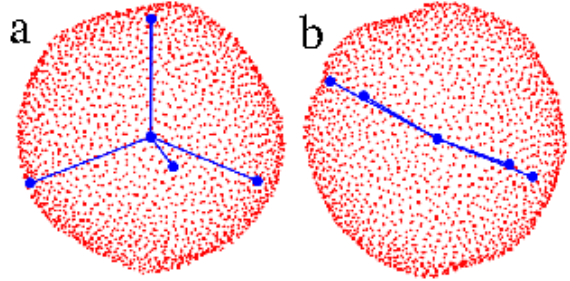}
\caption{ (color online) The simplest vesicle shape, at fixed volume, with only 
nonzero isotropic bending modulus $\kappa=20$ and weak nematic interaction 
$\epsilon _{LL}=1.5$. The four $s=+1/2$ defects (blue solid circles) fluctuate, at 
finite temperature, between (a) tetrahedral and (b) planar arrangements. The lines 
are the connectors between the centroid and the defects. The numerous small dots 
(red) show the vertices of the triangulated mesh outlining the surface of the vesicle.
Here the nematic interaction was chosen weak because at stronger interaction the free
energy barrier between (a) and (b) states will be higher which cannot be overcome 
only by thermal fluctuations. In the active case \cite{Bausch} active fluctuations 
make these states unstable.}
\label{fig.tetra}
\end{figure}

Although the dynamics of this active system is beyond the  scope of our equilibrium 
Monte-Carlo  simulation, some aspects of this system can still be understood from 
equilibrium physics of our model studied at physically relevant parameters.  Towards this 
we set $c_{\parallel}=0$ in our model, since MT does not have intrinsic curvature like FtsZ, 
and retain nonzero values of $k_{\parallel}, k_{\perp}, c_{\perp}$ and $\epsilon_{LL}$.   
It is well known that polymerization of actin and MT bundles, even in the absence of 
activity can give rise to tubular protrusions \cite{actinmembranePlos,libchaberbuckledMT}. 
That implies that membrane tubulation can occur at equilibrium also, however the dynamics 
of the tube and the vesicle has a purely nonequilibrium origin (motor acivity). The location of the 
tubes in Keber et. al's experiment \cite{Bausch} coincides with the locations of  +1/2 
nematic defects, where elastic strain in the nematic field is the highest.  However, tubulation
also requires extra area. So when the vesicle is tout (i.e., membrane tension high), the 
MT bundles cannot overcome the membrane tension to push out tubes, instead the bundles 
buckle.  This effect was seen in Ref\cite{libchaberbuckledMT}. For such an undeformed  
spherical vesicle the equilibrium positions of four +1/2 defects are the vertices of a 
symmetric tetrahedron, since +1/2 defects repel each other. Due to activity in 
Ref\cite{Bausch}, such an equilibrium state turns out to be unstable and 
the defects oscillate between a tetrahedral and a planar configuration \cite{Bausch}. 
Although, we cannot model this instability, our Monte-Carlo simulation at high bending 
rigidity ($\kappa$) and at high temperature shows that the defect positions indeed 
fluctuate between tetrahedron and planar configurations (see Fig.\ref{fig.tetra}) indicating 
closeness of these two states in the equilibrium energy landscape.  

Furthermore, when Keber et al \cite{Bausch} deflate the vesicle, by applying hypertonic stress,
it amounts to generating excess area (as the volume reduces) and as a result the vesicle deforms. 
To mimic this system we allow about $10\%$ excess area for the vesicle as before, 
but switch to relatively stronger nematic and bundling interactions: $\epsilon_{LL}=9$ and $c_{\perp}=1.2$. 
In addition, we set  $c_{\parallel}=0$ as MT filaments do not have intrinsic curvature.
When started from a sphere, the vesicle grows into an ellipsoid utilising the extra area. 
The four +1/2 defects arrange themselves into two pairs, one pair each migrating approximately 
to the opposite ends of the major axis. The strong nematic interaction ensures that more 
defects are not nucleated as defects possess high elastic energy. Subsequently four
tubes emerge from the four defects (Fig.6a). However as our system do not have any active 
dynamics the tubes do not change their positions. 
Interestingly, at these same parameter values, when we reduce the volume of the vesicle
further to one third of its value (with corresponding reduction in area) only two tubes emerged
(see Fig.6b). This occurred via merging of the defect pair into one +1 defect,  at each 
end of the major axis.  The corresponding size difference in the experimental figures 
are indicated in Fig.6c and d.

In our equilibrium model the filaments cannot slide/translate, unlike in the active case of 
Keber et al \cite{Bausch}, however the defects can move due to rearrangement of the nematics. 
As mentioned earlier, the areal density of nematics can be non-uniform in our model, 
because although there is one filament per vertex, the areal density of vertices is 
higher on the tubes than in the valleys. This could very well be the case in the 
experiment. For example, parallel arrangement of filaments on tubes, parallel to the
tube axis, will produce denser filament coverage and than that in the valleys. This is
consistent with higher fluorescent intensity from the tubes, reported in Osawa's 
experiments \cite{osawaEMBO}   
\begin{figure}[h]
\centering
\includegraphics[width=3in]{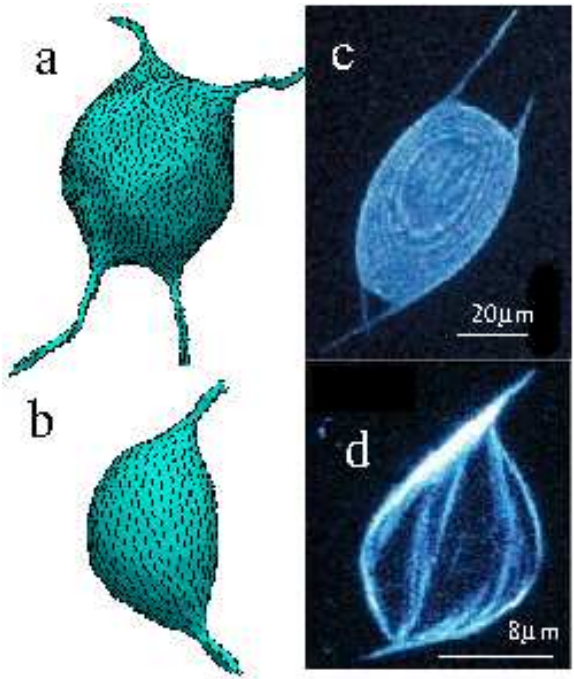}
\caption{ (color online) Effect of excess area: at fixed volume the area is fixed at $10\%$ 
excess over the corresponding sphere.  
(a) and (b) model partially deflated vesicles (c and d) of Ref\cite{Bausch} (with permission)  where 
initially spherical vesicles deformed into ellipsoidal shapes due to excess available area. 
(d) had lesser volume than (c), which resulted into half the number of tubes in (d). Model parameters
are the same for (a) and (b) :  $\kappa_0=0, \kappa_{\parallel}=25,\kappa_{\perp}=20, 
c_{\parallel}=0, c_{\perp}=1.2, \epsilon _{LL}=9$,  except that volume of (b) is 1/3-rd of (a). 
Despite presence of bundling effect $c_{\perp}$, strong nematic interaction allowed only finite 
number of defects (and hence tubes) as defects cause high elastic energy.}
\label{fig.baush}
\end{figure}

\begin{figure}[h]
\centering
\includegraphics[width=2.5in]{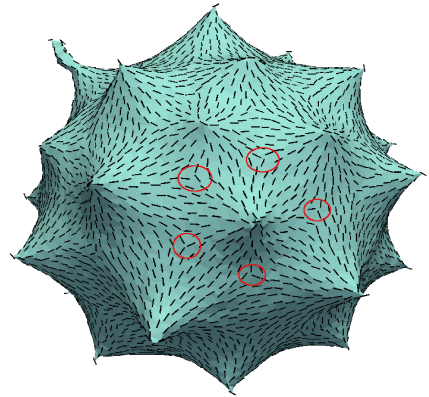}
\caption{ (color online) Arrangement of nematic defects on a tubulating vesicle. The red circles 
denote $s=-1/2$ defects which encircle the aster like $s=+1$ defects. The parameters 
are same as in Fig.~\ref{fig.tubes}. }
\label{fig.theorem}
\end{figure}

Nematic defects turn out to be hot spots of activity on the membrane. In our 
simulation tubes grow from sites of  nematic defects with a positive defect charge.
The charge $s$ of a nematic defect is defined as the total amount of rotation a nematic 
director undergoes as a closed loop is traversed around a defect.  
In Keber et. al. \cite{Bausch} defect sites on ellipsoid shaped vesicle constitute 
weak spots that encourage growth of MT bundles which in turn induce tubulation. 
Ladoux's group \cite{julia} has shown cell death and extrusion occurring predominantly
at the sites of $s=+1/2$, in nearly 2D epithelial tissue. 
It turns out that energetically it is favourable to co-localize nematic defects
and high membrane curvature. However in-plane arrangement of nematic defects on a 
closed surfaces must obey 
Poincare's index theorem (popularly known as the hairy ball theorem) which states
that the total topological charge ($s$) of all defects on a sphere must add up to 2 
(more generally to $2(1-g)$ on a 
surface of genus $g$) \cite{prost}. This constrains the number and location of the tubes on the 
vesicle surface. What precedes a tube is a defect of charge $s=1$ with  an aster 
like arrangement of nematics. 
The membrane deforms into a pointed structure (vertex) around the aster and produce  
a tube. The total positive charge increases with the number of vertices but 
this is efficiently nullified by the local arrangement of $s=-1/2$ defects around each vertex. 
Each vertex is surrounded by typically five or six,  and seldom seven, other vertices. 
Accordingly, the central vertex forms five, six or seven triangles, with the surrounding vertices.
Each triangle forms a concave valley and hosts a $s=-1/2$ defect 
(red circles in Fig.3). But each $s=-1/2$ defect is shared by three $s=1$ defects.
To compute total defect charge on the vesicle, we consider contributions from 
polygons formed by the $-1/2$ defects (red circles in Fig.~\ref{fig.theorem}). 
The net charge of a pentagon is $1 + 5\times(-1/2)\times (1/3)  = 1/6$,
while that for a hexagon is $1 + 6\times(-1/2)\times (1/3) = 0$, and heptagon is $-1/6$. 
One realisation of this pentagon-hexagon  arrangement  is the minimal soccer ball 
structure which has 12 pentagons, 20 hexagons and equivalently, thirty-two $+1$ and 
sixty $-1/2$ defects, with total charge adding up to 2. The simple picture that 
emerges is that each polygon, made of $s= -1/2$ defects, hosts one $s=+1$ defect (tube) 
at its centre. So in the framework of soccer ball structure the polygonal units are
formed by the $-1/2$ (valley) defects at the vertices.  In our initial guess we 
considered the hexagonal lattice dual to this where tubes were at the vertices
of the polygons. However, smaller and larger vesicles have different number of 
defects, for example, seventy-six $-1/2$ and forty $+1$ defects,  still adding up to  
2 (in fact Fig.~\ref{fig.theorem} has this structure).  The regular arrangement of valleys 
and tubes now can be matched with Figs.~\ref{fig.expt}f and ~\ref{fig.expt}g,  respectively. 
Note from Fig.~\ref{fig.theorem}, that each tube (+1 defect) typically has five to six  
nearest neighbours while each valley (-1/2 defect) has three nearest neighbours. 
Charge cancellation among nematic defects, on deformed axis-symmetric vesicles, 
have been discussed in Ref\cite{iglicSciRep2016}. 

In summary, we showed how filament bundle induced tubulation, can be modeled by 
introducing an induced anisotropic membrane curvature perpendicular to the 
filament's alignment.
We showed that, within our model, either of $c_{\parallel}$ and $c_{\perp}$  is
individually capable of causing tubulation. But the corresponding inclination 
of the nematic field on the tube surface is different in the two cases.  
Furthermore, when nematic interaction is weak (in case of FtsZ) the vesicle allows 
formation of many defects, and subsequently many tubes where  bundling interaction
leads to maximum energy gain. On the other hand for MT with strong nematic interaction 
only minimal number of defects and tubes form. Althouh GTP/ATP hydrolysis is common
to both cases (FtsZ/MT), we emphasize that the FtsZ case is not active in the 
conventional sense, because no physical filament movement is generated (in the 
absence of motors) unlike the case of kinesin driven MT case.  Although our model 
cannot address the active MT dynamics, our simulations indicates
that some of the features observed in the active MT induced vesicle shapes 
may have equilibrium origin. This includes emergence of the ellipsoidal
shape upon reduction of the vesicle volume and the influence of the vesicle volume
in determining the number of tubes (four versus two). 
The novel filament arrangement on the deformed vesicle surface is an interesting outcome 
of our numerical investigation and is difficult to predict a priori. This link between filament 
arrangement and vesicle shape may be useful for vesicle origami.
   



Gaurav kumar would like to acknowledge financial support from CSIR (India).


\begin{thebibliography}{10}

\bibitem{julichertube}
\newblock {I. Derenyi, F. Julicher, and J. Prost }.
\newblock {Physical review letters} {\bf 88}, 238101 (2002)

\bibitem{lasertweezer}
\newblock {F. Hochmuth, J.-Y. Shao, J. Dai, and M. P. Sheetz }.
\newblock {Biophysical journal} {\bf 70}, 358 (1996).

\bibitem{micropipette}
\newblock {R. Hochmuth, H. Wiles, E. Evans, and J. McCown }.
\newblock {Biophysical journal} {\bf 39},83 (1982). 

\bibitem{motor}
\newblock {A. Roux, G. Cappello, J. Cartaud, J. Prost, B. Goud and P. Bassereau }.
\newblock {Proceedings of the National Academy of Sciences} {\bf 99},5394 (2002). 

\bibitem{actinmembranePlos}
\newblock {Weichsel, J., and P. L. Geissler}.
\newblock { PLoS computational biology} {\bf 12 },e1004982 (2016). 

\bibitem{MesarecIglic2017}
\newblock {Mesarec, L., W. G ́o z  ́ d ́z, S. Kralj, M. Foˇsnariˇc, S. Peniˇc,
V. Kralj-Igliˇc, and A. Igliˇc,}.
\newblock {European Biophysics Journal } {\bf 46 }, 705-718  (2017). 

\bibitem{libchaberbuckledMT}
\newblock { Elbaum, M., D. K. Fygenson, and A. Libchaber,}.
\newblock { Physical Review Letters} {\bf 76  },  4078(1996).

\bibitem{Bausch}
\newblock {F. C. Keber, E. Loiseau, T. Sanchez, S. J. DeCamp, L. Giomi, M. J. Bowick, M. C. Marchetti, Z. Dogic, and A. R. Bausch}.
\newblock {Science} {\bf 345 },1135 (2014).


\bibitem{dynamin}
\newblock {Low, H. H., C. Sachse, L. A. Amos, and J. L ̈owe}.
\newblock {Cell} {\bf 139}, 1342–1352 (2009).

\bibitem{bundle}
\newblock {Hinshaw, J.}.
\newblock {Annual review of cell and developmental biology } {\bf  16}, 483–519 (2000).

\bibitem{ESCRT3}
\newblock {Adell, A. Y., D. Teis}.
\newblock {FEBS letters } {\bf  585}, 3191–3196 (2011).


\bibitem{peter2004bar}
\newblock { Peter, B. J., H. M. Kent, I. G. Mills, Y. Vallis, P. J. G. Butler,
P. R. Evans, and H. T. McMahon,}.
\newblock {Science } {\bf 303  }, 495–499 (2004).

\bibitem{ford2002curvature}
\newblock {Ford, M. G., I. G. Mills, B. J. Peter, Y. Vallis, G. J. Prae-
fcke, P. R. Evans, and H. T. McMahon, }.
\newblock {Nature } {\bf  }, 419:361 (2002).


\bibitem{zhao2013exo70}
\newblock { Zhao, Y., J. Liu, C. Yang, B. R. Capraro, T. Baumgart,
R. P. Bradley, N. Ramakrishnan, X. Xu, R. Radhakrishnan,T. Svitkina,}.
\newblock { Developmental cell} {\bf 26  },:266–278  (2013).


\bibitem{kozlov}
\newblock {Zimmerberg, J., and M. M. Kozlov }.
\newblock {Nature reviews Molecular
cell biology } {\bf  }, 7:9 (2006).


\bibitem{Dimova}
\newblock {Y. Li, R. Lipowsky, and R. Dimova}.
\newblock {Proceedings of the National Academy of Sciences} {\bf 108 },4731 (2011).


\bibitem{lipidtilt}
\newblock {Schnur, J. M. }.
\newblock {Science } {\bf262  }, :1669–1676 (1993).


\bibitem{osawaEMBO}
\newblock {M. Osawa, D. E. Anderson, and H. P. Erickson}.
\newblock {The EMBO journal} {\bf 28},3476 (2009).

\bibitem{bundle}
\newblock {Erickson, H. P., D. W. Taylor, K. A. Taylor, and D. Bramhill}.
\newblock { Proceedings of the National Academy of
Sciences} {\bf 93 }, 519–523 (1996).

\bibitem{Sci17Huang}
\newblock {X. Yang, Z. Lyu, A. Miguel, R. McQuillen, K. C. Huang and J. Xiao}.
\newblock {Science} {\bf 355 },744 (2017).


\bibitem{Sci17garner}
\newblock {A. W. Bisson-Filho, Y.-P. Hsu, G. R. Squyres, E. Kuru, F. Wu, C. Jukes, Y. Sun, C. Dekker, S. Holden, M. S. VanNieuwenhze, et al.}.
\newblock {Science} {\bf 355},739 (2017).

\bibitem{zringprl}
\newblock {B. Ghosh and A. Sain}.
\newblock {Physical review letters} {\bf 101},178101 (2008).

\bibitem{osawaSci}
\newblock {M. Osawa, D. E. Anderson, and H. P. Erickson}.
\newblock {Science} {\bf 320 },792 (2008).

\bibitem{canham70}
\newblock {Canham, P. B.}.
\newblock {Journal of theoretical biology } {\bf 26 },61–81  (1970).


\bibitem{helfrich73}
\newblock {Helfrich, W }.
\newblock {Z. Naturforsch. C } {\bf 81 }, :041922 (1973).



\bibitem{SunilPRE}
\newblock {N. Ramakrishnan, P. S. Kumar, and J. H. Ipsen}.
\newblock {Physical Review E} {\bf 81},041922 (2010).


\bibitem{Sunilbiopj}
\newblock {N. Ramakrishnan, P. S. Kumar, and J. H. Ipsen}.
\newblock {Biophysical journal} {\bf 104 },1018 (2013).

\bibitem{SunilMacromolecule}
\newblock {N. Ramakrishnan, P. Kumar, and J. H. Ipsen}.
\newblock {Macromolecular Theory and Simulations} {\bf 20 },446 (2011).


\bibitem{prost}
\newblock {Lubensky, T., and J. Prost }.
\newblock { Journal de Physique II} {\bf  2}, :371–382 ( 1992 ).

\bibitem{nelson}
\newblock {Nelson, D. R }.
\newblock {Nano Letters } {\bf 2 }, :1125–1129  ( 2002).

\bibitem{bowickPRL}
\newblock { Shin, H., M. J. Bowick, and X. Xing}.
\newblock {Physical review letters } {\bf 101  }, :037802 ( 2008 ).


\bibitem{Chaikin}
\newblock { Chaikin, P. M., and T. C. Lubensky,}.
\newblock {Cambridge university press } {\bf  }, : ( 2000 ).

\bibitem{LL}
\newblock {P. Lebwohl}.
\newblock {Phys. Rev. A} {\bf 6 },426 (1972).



\bibitem{iglic2005}
\newblock {Igliˇc, A., B. Babnik, U. Gimsa, and V. Kralj-Igliˇc}.
\newblock { Journal of Physics A:
Mathematical and General} {\bf38  }, :8527 ( 2005 ).



\bibitem{AnisotropicMemCurv2016simul}
\newblock {G ́omez-Llobregat, J., F. El ́ıas-Wolff, and M. Lind ́en }.
\newblock {Biophysical journal } {\bf 110 },197–204  (2016  ).


\bibitem{Iglic2017}
\newblock {Mesarec, L., A. Igliˇc, and S. Kralj, }.
\newblock {Adv Biomed Res Innov} {\bf  },4:2. ( 2018).

\bibitem{wortis}
\newblock {HW, G. L., M. Wortis, and R. Mukhopadhyay,}.
\newblock {Proceedings of the National Academy of Sciences} {\bf99  },16766–16769 (2002 ).


\bibitem{onsphere}
\newblock {Zhang, R., Y. Zhou, M. Rahimi, and J. J. De Pablo }.
\newblock {Nature
communications } {\bf 13 },13483 (2016 ).

\bibitem{voigt}
\newblock {F. Alaimo, C. Kohler, and A. Voigt}.
\newblock {arXiv preprintarXiv:} {\bf }1703.03707 (2017).

\bibitem{julia}
\newblock {T. B. Saw, A. Doostmohammadi, V. Nier, L. Kocgozlu, S. Thampi, Y. Toyama, P. Marcq, C. T. Lim, J. M. Yeomans, and B. Ladoux}.
\newblock {Nature} {\bf 544 },212 (2017).


\bibitem{iglicSciRep2016}
\newblock { Mesarec, L., W. G ́o z  ́ d ́z, A. Igliˇc, and S. Kralj}.
\newblock {Scientific reports } {\bf 6 },27117. (2016 ).


\end{thebibliography}
\end{document}